\documentclass{aa}  

\usepackage{xcolor}
\usepackage{colortbl}
\usepackage{amsmath}
\usepackage{multirow}
\usepackage{makecell}
\usepackage{caption}
\usepackage{adjustbox}
\usepackage{float}
\usepackage{diagbox}
\usepackage{lipsum}
\usepackage{hyperref}
\usepackage{selinput}
\usepackage{nicematrix}
\usepackage{tikz}
\usetikzlibrary{calc,matrix,shapes.callouts}
\usepackage{txfonts}
\usepackage{graphicx}

\newcommand{\Lsd}{L_{\rm sd}}

\newcommand{\mcdot}{\dot m_{\rm c}}

\newcommand{\vpsr}{v_{\rm psr, \, w}}
\newcommand{\vc}{v_{\rm c, \, w}}
\newcommand{\aib}{a_{\rm IB}}
\newcommand{\rsh}{r_{\rm sh}}
\newcommand{\rcrit}{r_{\rm t,crit}}

\newcommand{\betaiso}{\beta} 
\newcommand{\rt}{r_{\rm t}}
\newcommand{\mpsrdot}{\dot m_{\rm psr,w}}

\begin{document} 

   \title{ Hydrodynamical simulations of wind interaction in spider systems }
   \subtitle{A step toward understanding transitional millisecond pulsars }

   \author{C. Guerra
          \inst{1}
          \and
          Z. Meliani\inst{1}
          \and
          G. Voisin\inst{1}
          }

   \institute{Laboratoire Univers et Théories, Observatoire de Paris, Université PSL, Université de Paris cité, CNRS, F-92190 Meudon, France \\
    \email{corentin.guerra@obspm.fr ; zakaria.meliani@obspm.fr ; guillaume.voisin@obspm.fr}
             }

   \date{Received Mai 07, 2024; accepted July 11, 2024}

\abstract
{The detected population of "spiders", referring to millisecond pulsar binaries, has significantly grown in the past decade thanks to multiwavelength follow-up investigations of unidentified \textit{Fermi} sources. These systems consist of low-mass stellar companions orbiting rotation-powered millisecond pulsars in short periods of a few hours up to day. Among them, a subset of intriguing objects called transitional millisecond pulsars (tMSPs) has been shown to exhibit a remarkable behavior, transitioning between pulsar-binary and faint low-mass X-ray binary states over a span of a few years. 
}
   {Our objective is to study the interaction of stellar winds in tMSPs in order to understand their observational properties. To this end we focus on the parameter range that places the system near Roche-lobe overflow.}
   {Employing the adaptative mesh refinement (AMR) AMRVAC 2.0 code, we performed 2D hydrodynamical (HD) simulations of the interaction between the flows from both stars, accounting for the effects of gravity and orbital motion. }
   {By studying the mass loss and launch speed of the winds, we successfully recreated two phenomenologically distinct regimes: the accretion stream and the pulsar radio state. We also identified the tipping point that marks the sharp transition between these two states. In the accretion stream state, we discover a very strong variability induced by the pulsar wind. In the pulsar state, we reconstructed the corresponding X-ray light curves of the system that produces the characteristic double-peak pattern of these systems. The position of the peaks is shifted due to orbital motion and the leading peak is weaker due to eclipsing by the companion.}
   {This study highlights the importance of gravity and orbital motion in the interaction between the companion and pulsar winds.
   Our setup allows the study of the complex interaction between the pulsar wind and an accretion stream during mass transfer.
   We suggest that a smaller leading peak in X-rays is indicative of a nearly edge-on system.}

    
   \maketitle
%
\section{Introduction}
Among the pulsar population, millisecond pulsars (MSPs) are characterized by their weaker surface magnetic fields ($B_{\rm psr} \sim 10^8 - 10^9 \mathrm{G}$), faster spin periods ($P_{\rm spin} \sim 10^{-3} \mathrm{s}$), and smaller spin-down rates ($\Dot{P}_{\rm spin}\sim 10^{-20}  \mathrm{s \cdot s^{-1}}$). The recycling scenario postulates that mass accretion from a binary companion \citep{alpar1982new} can lead to their spin-up to millisecond spin periods. Indeed, this scenario is supported by the fact that MSPs are commonly found in a binary system \citep{manchester2017millisecond}.

In recent years, significant progress has been made in understanding binary systems known as "redbacks" (RBs) and "black widows" (BWs), which are also referred to as spider pulsars \citep{Eichler_Levinson_1988ApJ...335L..67E,Roberts_2013IAUS..291..127R}. These systems consist of rotation-powered  millisecond pulsars that reside in compact binary orbits. 
Indeed the orbital period of these systems is typically $P_{\rm orb} \le $ 1 \rm{day} \citep{Chen_etla_2013ApJ...775...27C,breton2013discovery}, and the two components are separated by $a_{\rm IB} \sim 10^{11}$ \rm{cm}.
The companion is a degenerate or semi-degenerate star of approximately 0.01 to 0.05 solar masses ($M_\odot$)  for BWs, and about 0.1 to 0.5 $M_\odot$ for RBs \citep{Chen_etla_2013ApJ...775...27C}. The first black widow pulsar discovered was PSR B1957+20 \citep{Fruchter_etal_1988Natur.333..237F}, and the first redbacks discovered were PSR J0024-7204W \citep{bogdanov2005x} and PSR J1023+0038 \citep{Archibald_etal_2009Sci...324.1411A}.
The term "spider systems" evokes the possible ablation of the companion star \citep{phinney1988ablating, ginzburg2020black} due to the impact of the pulsar wind. Additionally, the notion that these systems could lead to the complete ablation of the companion star may serve as the missing link connecting binary pulsar systems to isolated millisecond pulsars.

The discovery of  several dozen spider systems over the last decade has been made possible through follow-up observations of \textit{Fermi}/LAT unidentified sources \citep[e.g.,][]{li2018multiwavelength, strader2019optical,clark21}. 
The effort involved in searching for spider systems has been motivated by indications that they may harbor massive neutron stars beyond $2M_\odot$ \citep[e.g.,][]{linares2019super}, although it has been shown that these measurements can suffer from large systematic errors \citep[e.g.,][]{voisin2020model,romani2021psr,clark2023neutron}.
These comprehensive multiwavelength observations have provided valuable insights into their properties and characteristics. 
The radio light curves show a pulsar eclipse. This could be caused by plasma ejected by the companion star \citep[e.g.,][]{polzin18}. 

The recent emergence of a  subclass of millisecond pulsars (MSPs), known as transitional millisecond pulsars (tMSPs) (PSR J1023+0038 \citep{Archibald_etal_2009Sci...324.1411A}, PSR J1824-2452I \citep{Papitto_etal_2013Natur.501..517P}, PSR J1227-4853 \citep{bassa2014state, roy2015discovery}) has posed significant challenges with respect to our understanding of the physical processes governing the evolution of these astrophysical objects. These systems undergo transitions between the radio-pulsar and accretion states on timescales much shorter than typical stellar evolution timescales.

The first discovered tMSP was PSR J1023+0038 \citep{Archibald_etal_2009Sci...324.1411A}. Initially, optical observations in 2001 misclassified the previously known radio source as a cataclysmic variable with an accretion disk around the pulsar, based on the detection of double-peaked emission lines \citep{bond2002first, szkody2003cataclysmic}. However, in 2004, the emission lines had disappeared, leading to the subsequent identification and categorization of the source as a radio pulsar \citep{thorstensen2005first, Archibald_etal_2009Sci...324.1411A}. PSR J1023+0038 \citep{Papitto_Torres_2015ApJ...807...33P} remained in the radio pulsar state until 2013 when the radio pulsations ceased, and there was a significant increase in optical/UV, X-ray, and $\gamma$-ray activity, indicating a reactivation of accretion on the pulsar. The transition took place within less than two weeks \citep{stappers2014state} and the system has since remained in this state.

Multiwavelength observations from radio to gamma rays have the potential to constrain every component of the system. Pulsar timing provides accurate ephemeris of the orbital motion up to a degeneracy between masses and the inclination angle of the system. This degeneracy can be lifted thanks to the modeling of the optical light curve of the companion, leading to mass measurements which have shown that these neutron stars tend to be rather massive \citep[e.g.,][]{clark2023neutron,linares2019super,van2011evidence}. Pulsed X-ray emission was also detected during the accretion state of PSR J1023+0038, allowing for the evolution of the rotational phase during accretion to be tracked \citep{jaodand2016timing}.

Unpulsed X-ray emissions revealed peaks of emissions near pulsar inferior or superior conjunctions, depending on the system, which have been interpreted as the signature of an intrabinary shock (IBS) wrapping around one or the other star depending on the balance between the two winds \citep{Sanchez_Romani_2017ApJ...845...42S,wadiasingh2017constraining,gentile2014x,kandel2021xmm,papitto2021transitional}. Indeed, the tangential discontinuity separating the companion and the pulsar wind is an efficient site of particle acceleration \citep{harding1990acceleration,arons1993high,cortes2022global} as the shocked material from the pulsar wind can reach relativistic bulk velocities leading to a non-thermal emission. 
Synchrotron radiation (SR) and inverse compton (IC) are two processes responsible for the emission of these high-energy photons. 
In the accretion state of tMSPs, the disk is characterized by flaring light curves in optical and X-ray \citep{bogdanov2015coordinated,linares2022x}.  In the case of PSR J1023+0038, the onset of the last accretion state has been accompanied by a fivefold increase in unpulsed gamma-ray flux the origin of which is yet unclear \citep{stappers2014state}. The unique behavior of tMSPs together with this consistent set of multi-wavelength observations offers an opportunity to study wind-wind interactions and accretion physics in the same system, on top of providing strong evidence in support of the recycling scenario.
The various states observed in spider systems are undoubtedly influenced by the balance between the ram pressure exerted by the pulsar wind on the matter lost by the companion. The interplay of gravitational and rotational forces significantly shapes the flow of stellar matter \citep{wadiasingh2018pressure}. 

Two main scenarios have been studied: the case of an accreting neutron star and the case where the pulsar is active, generating a wind that interacts with the wind of the companion that results in a shock surrounding the companion star \citep{Benvenuto_etal_2015ApJ...798...44B, Sanchez_Romani_2017ApJ...845...42S}. The interplay between these forces drives the observed transitions and underlies the complex behavior exhibited by tMSPs.
Until now, the numerical investigation of the interaction between the wind of a companion star and a pulsar wind in a binary system has predominantly focused on systems involving a massive star. These studies have employed non-relativistic smoothed particle hydrodynamic (SPH) simulations in 3D \citep{Romero_etal_2007A&A...474...15R}, as well as classical and relativistic hydrodynamic simulations in 2D \citep{bogovalov2012modelling, Lamberts_etal_2013A&A...560A..79L}, where  the orbital motion and considered axisymmetric cases  were neglected \citep{Paredes-Fortuny_etal_2015A&A...574A..77P}. Recent advances include 3D relativistic simulations, including the effects of orbital motion, conducted by \citep{Bosch-Ramon_etal_2015A&A...577A..89B,Huber_etal_2021A&A...649A..71H}.


In this paper, we employ hydrodynamical numerical simulations to investigate the wind-wind interaction in 2D, at the equatorial plane between the pulsar wind and the wind from the low-mass companion star. Bearing in mind the transitional nature of tMSPs, we focus on the region of the parameter space surrounding the limit between accreting and non-accreting neutron star (NS) behavior. Additionally, we model the X-ray light curve in the IBS regime.

The paper is structured as follows. In Sect.~\ref{Sec:numerical_model} we describe the physics of the numerical model we used in our study. This includes the description of the numerical model of the pulsar and companion wind. 
In Sect. \ref{sec:ibs_model}, we summarize the IBS modelization, as well as the model of SR emission used to compute the associated X-ray light curves.
The simulation setup is described in Sect. \ref{sec:simulations_setup}. Finally, we discuss the outcomes of our study in Sect. \ref{sec:results} and present the application for the observations of the pulsar spider systems in Sect.~\ref{sec:xLC}.

\section{Physical and numerical model}\label{Sec:numerical_model}
To construct a comprehensive model of spider systems, it is essential to explore various physical processes, phenomena and interactions involved in the system. These include the irradiation of the donor star's atmosphere by the pulsar, the transfer of gas from the companion star and the subsequent formation of the accretion disc, as well as the wind-wind interaction between theflow from the companion and the pulsar wind. By investigating these aspects, we aim to develop a coherent understanding of the complex dynamics at play in the system.
\begin{figure}
\begin{center}
\includegraphics[width=\columnwidth]{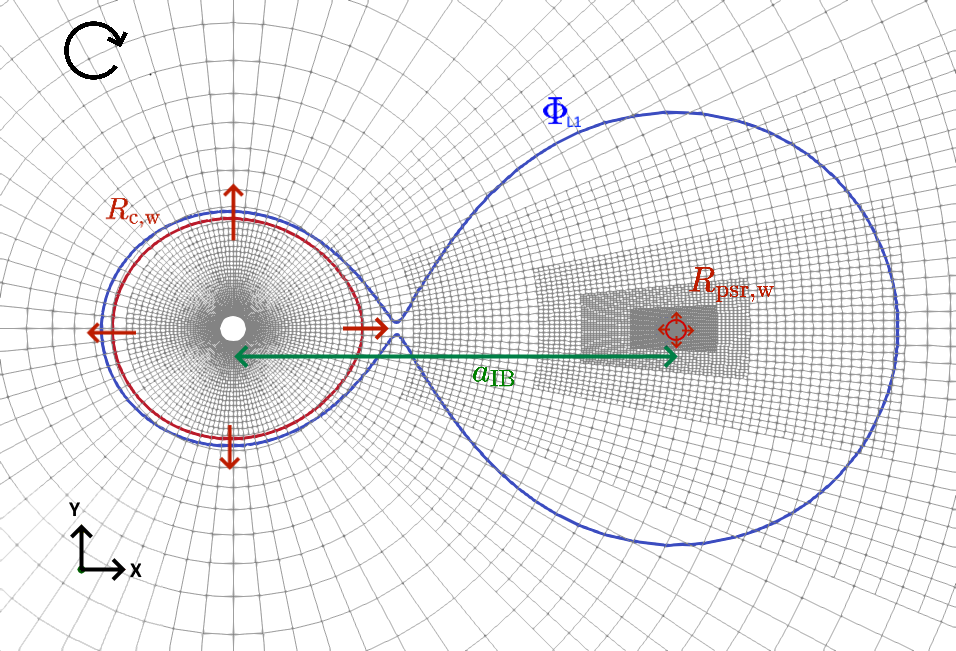}
\caption{Sketch of the grid setup used in simulations. 
The companion star is at the center of the cylindrical grid, and the pulsar at a distance $\aib$ along X axis. Wind injection surfaces are shown in red, gravitational potential of Lagrange 1 point is shown in blue. Top left arrow indicates sens of rotation.}
\label{sketch}
\end{center}
\end{figure}

\begin{table*}
        \centering
        \caption{
        Main parameters in use}
        \begin{tabular}{lc}
        \hline
        Parameters &  Values \\
        \hline
        \hline
    Binary separation distance $a_{\rm IB}$ (cm) & $10^{11}$\\
    Binary period   $P$  (s)  &  $1.28\times 10^{4}$\\ 
    Binary center of mass distance from companion star $r_{\rm center}$ (cm)  &  $0.78\times 10^{11}$\\
    Binary orbital speed $v_{\rm orb}$ (km/s)& $490$\\
    \hline
    Pulsar mass    M$_{\rm psr}$ (M$_{\odot}$)      &  $1.4$   \\
    Pulsar wind spin down luminosity $L_{\rm sd}$ ($\rm erg.s^{-1}$) & $ 10^{35}$\\
    Pulsar wind mechanical luminosity in sim. $L_{\rm psr,w}$ ($\rm erg.s^{-1}$) & $5 \times 10^{31}$\\
    Pulsar wind initial speed $\vpsr$ (km/s) & $1\times 10^{4}$\\
    Pulsar wind mass flux $\dot{m}_{\rm psr,w}$ ($M_{\odot}/yr$) & $1.5\times10^{-12}$\\ 
    Pulsar wind surface radius $R_{\rm psr,w}$ (cm) & $2.5\times 10^{9}$\\  
    \hline
    Companion mass   M$_{\rm c}$ (M$_{\odot}$)  & $0.4$\\
    Companion Roche lobe radius R$_{\rm c}$ (cm) & $0.37 \times 10^{11}$ \\
    Companion wind mechanical luminosity $L_{\rm c}$ ($ \rm erg.s^{-1}$) & $ 5\times 10^{29} - 1\times 10^{31}$ \\
    Companion wind initial speed $\vc$ ($ \rm km/s$) & $60 - 200$ \\
    Companion wind mass flux $\dot{m}_{\rm c,w}$ ($\rm M_{\odot}/yr$) & $4.5\times 10^{-10}-9\times 10^{-10}$ \\
    Companion wind surface radius $R_{\rm c, w}$ at $y=0$ (cm) & $0.3 \times 10^{11}$\\
    Companion wind surface temperature $T_{\rm c}$ (K) & 4480\\
    \hline
    Constraints &   \\
    \hline
    \hline
    Companion wind max initial speed $v_{\rm c,max}$ ($\rm km/s$) & 0.03 $\vpsr$ = $ 300$ \\
    Companion wind max mass flux $\dot{m}_{\rm c,max}$ ($ \rm M_{\odot}/yr$) & $1\times 10^{-8}$ \\
    Companion wind max mechanical luminosity $L_{\rm c,max}$ ($\rm erg.s^{-1}$) & $3\times 10^{32}$ \\

\hline\\ 
        \end{tabular}
\label{tab:paper_parameters}
\end{table*}

\subsection{Orbital configuration}
In this study, we chose to investigate the case of a RB system prone to a transitional behavior, consisting of a pulsar with a mass of $M_{\rm psr} = 1.4 M_{\odot}$ and a companion star with a mass of $M_{\rm c} = 0.4 M_{\odot}$.

The separation distance between the pulsar and the companion star is $a_{\rm IB} = 10^{11}$ cm, resulting in a revolution time of  $P=\sqrt{\frac{4\pi^2a_{\rm IB}^3}{G\left(M_{\rm psr}+M_{\rm c}\right)}} \approx 1.28\times 10^{4}$ s (equivalent to 3.56 hours), where $G$ is the gravitational constant.  Additionally, the center of mass of the system is located at a distance of $r_{\rm center} \approx 0.78 a_{\rm IB}$ from the companion star.

In our analysis, we assume a circular orbit, following the common circularization pattern observed in spider systems. It is worth noting that in such systems, the companion star is typically tidally locked \citep{Hurley_etal_2002MNRAS.329..897H}; however, the eccentricity is usually very small \citep[e.g.,][]{Voisin_etal_2020MNRAS.492.1550V}.

\subsection{Winds}

We consider the fact that the winds from the pulsar and the companion are non-magnetized, isotropic, supersonic, and originate from a surface surrounding each one of them.  The relation between the wind mechanical luminosity $L_{w}$, mass flux $\dot{m}_{w}$, and speed $v_{w}$ is:
\begin{equation}
    \label{Lmv}
     L_{w} = \dot{m}_{w} v_{w}^2 / 2 \, .
\end{equation}

At the first level, the interaction between the pulsar wind and the companion wind can be characterized by the ratio of their momentum fluxes \citep{canto1996exact,parkin20083d,bogovalov2012modelling, Sanchez_Romani_2017ApJ...845...42S}, given by :
\begin{equation}  
\label{Eq:ratio_beta}
\beta =    \frac{\dot{\Pi}_{\rm c}}{\dot{\Pi}_{\rm psr} } = \frac{\dot{m}_{\rm c,w} \vc}{\dot{m}_{\rm psr,w} \vpsr}\,,
\end{equation}
where $\dot{m}_{\rm c,w}$ and $\vc$ are the mass flux and speed of the companion wind, and $\dot{m}_{\rm psr,w}$ and $\vpsr$ are the mass flux and speed of the pulsar wind.

The influence of the orbital properties on the wind from the companion star and on the wind from the NS can be characterized by two dimensionless quantities. Both are given at the surface where the wind is set. 
The first expresses the binding of the flow to the system, given by the ratio between wind speed and the escape speed $v_{\rm esc}$ as follows: 

\begin{equation}\label{Eq:vgrav_ratio}
\xi_{\rm grav} = \frac{v_{\rm w}}{v_{\rm esc}} \,.
\end{equation}

The second expresses the influence of the rotation of the system, specifically the Coriolis force. It is determined by the ratio of the wind speed to the rotation speed $v_{\rm rot}\sim 490$ km/s :

\begin{equation}\label{Eq:vrot_ratio}
\xi_{\rm rot} = \frac{v_{\rm w}}{v_{\rm rot}} = v_{\rm w} \left ( \sqrt{\frac{G\,\left(M_{\rm psr}+M_{\rm c}\right)}{a_{\rm IB}}} \right )^{-1}\,.
\end{equation}

\subsubsection{Pulsar wind}\label{psr wind}

The supersonic pulsar wind is emitted from the surface of a sphere with a radius of $R_{\rm psr,w}=0.025 a_{\rm IB}$, which surrounds the pulsar, as demonstrated by the model outlined in \citep{meyer2022rectangular}.


Throughout this paper, we consider a pulsar spin-down luminosity of $\Lsd = 10^{35}$ erg/s, which is typical of known tMSPs, where a fraction $(1-b)=0.1$ accounts for the radiated luminosity in $\gamma$-rays. The radiative luminosity in the hard X-ray band is assumed to be negligible compared to the $\gamma$-ray luminosity.
The speed of the pulsar wind is set (as seen in Table \ref{tab:pulsar_parameters}) so that the physical hierarchy of the system is respected, that is, the effects of gravity and rotation on the pulsar wind are nearly negligible.

\begin{table}[h]
    \centering
    \caption{Mechanical luminosity (Eq. \ref{Lmv}), initial speed, ratio of the speed with the orbital speed and escape speed, and Mach number of the pulsar wind.}
    \begin{tabular}{|c|c|c|c|c|}
        $L_{\rm psr, w}$ (erg/s) & $\vpsr$ (km/s)& $\xi_{\rm psr,grav}$ & $\xi_{\rm psr,rot}$ & $\mathcal{M}_{\rm psr,w}$ \\
        \hline
       $ 5\times 10^{31}$ & $1\times 10^{4}$ & $2.6$ & $20$ &$50$ 
    \end{tabular}
    \label{tab:pulsar_parameters}
\end{table}

However it's important to acknowledge that this speed is significantly lower than the realistic pulsar wind speeds, which can reach a Lorentz factor of $10^6$ \citep[e.g.,][]{kennel1984confinement}. The choice of a reduced pulsar wind speed can lead to significant changes in its properties, affecting compression rates, velocities, and consequently the positioning of shocks. It can also affect the development of associated instabilities. 
Since the momentum ratio is expected to primarily determine the behavior of the interaction between the two winds \citep{wilkin1996exact}, the mass flux of the wind $ \dot{m}_{\rm psr,w}$ is set such that momentum flux of a relativistic wind with a mechanical luminosity $b\Lsd$ is reproduced, following  :

\begin{equation}
\dot{\Pi}_{\rm psr}  = \dot{m}_{\rm psr,w} \vpsr = b\Lsd/c .
\end{equation} 
We consider that this setup already provides a valuable analog to the actual system at a significantly reduced computational cost compared to a relativistic version. 
The pulsar wind in our simuations is set with a mechanical luminosity $L_{\rm psr,w}=\dot{m}_{\rm psr,w} \vpsr^2/2$ following Eq. \ref{Lmv}. In the 2D approach the mass flux at a distance $r$ is : $\mpsrdot(r) =  b\Lsd / c\vpsr =  \mpsrdot(R_{\rm psr,w})\, r / R_{\rm psr,w}$. To reproduce correct momentum flux at a distance $r=\aib$, the initial mass flux of the wind, $\mpsrdot(R_{\rm psr,w}) = 1.5 \times 10^{-12} \mathrm{M_{\odot}/yr}$, setting $L_{\rm psr,w}=5\times 10^{31} \rm erg/s$.

In addition we make the assumption that all Poynting flux has been converted into kinetic flux \footnote{We note that the kinetic luminosity of the pulsar wind evolves with time $t$, resulting from the pulsar spin-down \citet{Lorimer_Kramer_2004hpa..book.....L}. However, the timescale given by the orbital period $P$ is ranging from a few hours up to a day, while the pulsar wind power decays over a characteristic timescale of $\tau_{\rm psr}\approx 1$ Gyr. Therefore, the pulsar spin-down power is considered constant during the simulation.}. 
However, it is important to note that a realistic pulsar wind is expected to be strongly asymmetric. This asymmetry arises from the angular distribution of the Poynting flux, which follows a $\sin^2{\theta}$ profile according to the analytical model \citep{Michel_1973ApJ...180L.133M}, and even a $\sin^4{\theta}$ profile based on numerical simulations \citep{Tchekhovskoy_etal_2013MNRAS.435L...1T}, where $\theta$ represents the polar angle.


\subsubsection{Companion wind}\label{Subsec:companion_wind}

In this work, we explore the effect of varying the two parameters characterizing the companion wind. The mass-loss rate is constrained by stellar evolution considerations to remain below $\dot m_{\rm c, w, max} = 10^{-8} \rm M_\odot /yr$ \citep{Chen_etla_2013ApJ...775...27C, Benvenuto_etal_2015ApJ...798...44B}. In order to reproduce the correct hierarchy of speeds between the two winds we set $\vc < v_{\rm c,max}$ = $300$ km/s = 0.03 $\vpsr$. 
This limit is not very constraining, as larger speeds would cause the matter to escape in all directions instead of being constrained to flow through $\rm L_1$.

The wind of the companion is characterized by a Mach number $\mathcal{M}_{\rm c, w}$ such that the surface temperature of the companion is $T_{\rm c} \sim 4480 $ K, typical value for RB systems. We explore distinct scenarios by varying the initial speed $\vc$ (with values between $60$ and $200$ km/s) and the mass flux of the wind $\mcdot$ (with values between $4.5 \times 10^{-10}$ $\rm M_{\odot}/yr$ and $ 9 \times10^{-10}$ $\rm M_{\odot}/yr$). 
This is summarized in Table \ref{tab:stellar_wind_parameters}.


The wind of the companion is set with sufficient specific kinetic energy to overcome the gravitational potential barrier to the Lagrange point $\rm L_1$. This criterion is satisfied by considering the conservation of Bernoulli energy along the flow streamline, given by:
\begin{equation}\label{Eq:companion_wind_bernoulie}
\xi = \frac{1}{2} v_{\rm c, w}^2 +
\frac{c_{s, w}^2}{\gamma - 1}\,+\,\Phi\,,
\end{equation}
where $\gamma$ represents the polytropic index, $v_{\rm c, w}$ denotes the wind speed of the companion in the rotating frame, and  $c_{s, w}= v_{\rm c, w}/\mathcal{M}$ the sound speed of the companion wind.  Lastly, $\Phi$ represents the effective gravitational energy in the rotating frame, accounting for the gravitational potential energy of the system, it is given in case of synchronous rotation by :
\begin{equation}\label{Eq:companion_wind_gravitation_potentiel}
\Phi\,=-\left [\,\frac{G\,M_{\rm psr}}{r_{\rm psr}}+\frac{G\,M_{\rm c}}{r_{\rm c}}+\frac{1}{2}\,\frac{G \left(M_{\rm psr}+M_{\rm c}\right)}{a_{\rm IB}^3}\,r_{\rm center}^2\ \right ]\,,
\end{equation}
where $r_{\rm psr}$, $r_{\rm c}$ and $r_{\rm center} $ are respectively the distance from pulsar, the companion star and the center of mass.

The fluid elements at the surface of the companion star that can escape the gravitational potential and cross the Lagrange point $\rm L_1$ must have a minimal specific kinetic energy. Thus, the Bernoulli energy at the Lagrange point, $\rm L_1$, is then $\xi=\Phi$. As a consequence, the wind speed at the surface of the companion star must exceed the escape speed given by :

\begin{equation}\label{Eq:companion_wind_speed_min}
v_{\rm c, min} = \left[\frac{\left(\Phi_{\rm L1}-\Phi_{\rm c, surface}\right)}{\frac{1}{2} + \frac{\mathcal{M}^{-2}}{\gamma-1}}\right]^{\frac{1}{2}}\,,
\end{equation}
where $\Phi_{\rm L1}$ represents the gravitational potential at the Lagrange point $\rm L_1$, $\mathcal{M}$ the mach number of the companion wind at its surface, and $\Phi_{\rm c, surface}$  the gravitational potential at the surface of the companion star. 

In this paper, the wind emanates from the equipotential surface with an effective potential $\Phi_{\rm c,surface}=\Phi(x=R_{\rm c,w},y=0)$, ((x, y) coordinates are defined in Fig. \ref{sketch}).

The influence of the gravity on the companion star flow is then given by the ratio between wind speed and the threshold speed necessary for the wind to fill the Roche Lobe and cross the Lagrange point $\rm L_1$ ( Eq. \ref{Eq:companion_wind_speed_min}): 
\begin{equation}\label{Eq:vmin_ratio}
\xi_{\rm c, grav} = \frac{v_{\rm c,\, w}}{v_{\rm c,min}} \,.
\end{equation}
The wind is set with speed such that mass transfer occurs through Roche-lobe overflow (RLOF), so the $\xi_{\rm c,\, grav}$ is set to be on the order of 1. As a result, the dimensionless rotational parameter $\xi_{\rm c,\, rot}$ is about one-third, so the rotation is expected to play a significant role.  

We define the Roche lobe filling ratio based on the position of the surface along the line of center, $R_{\rm c, w}=0.3 a_{\rm IB}$, which corresponds to about $80\%$ of the Roche lobe along this direction. 
Indeed, as the companion star is typically irradiated by the pulsar wind, this leads to its expansion, resulting in the filling of a significant portion of its Roche lobe, exceeding $90\%$ \citep[e.g.,][]{Crawford_etal_2013ApJ...776...20C} and also exceeding the radius of an isolated main-sequence star with a similar mass. 
This behavior is consistent with a quasi-RLOF state, facilitating mass and angular momentum transfer from the companion star to the pulsar \citep{Benvenuto_etal_2015ApJ...798...44B}. 

Furthermore, the companion star is tidally locked to the pulsar \citep{phinney1988ablating}. 
Combined with irradiation, this results in a potentially strong anisotropy of the surface temperature of the companion with a hotter "day" side.  
However, for the purposes of this initial study, our main focus is on the effect of wind-wind interaction. Therefore, we do not consider this effect in the cases presented in this paper.

Our current HD approach does not allow us to capture the dynamics associated with the potential companion magnetic field. Some studies have suggested that it may be an important element in high-energy emission scenarios \citep{sim2024} or in irradiation patterns \citep{Sanchez_Romani_2017ApJ...845...42S}.

\subsection{Heating of the companion star's wind} \label{subsec:heat}
The observed high effective surface temperature of the companion star may be influenced by various factors. These factors are likely to act in a complementary manner. One proposed mechanism is the irradiation by high-energy pulsar photons,  
considering that a fraction $(1-b)$ of the pulsar's spin-down luminosity $\Lsd$ is deposited at the surface of the companion star \citep{Benvenuto_etal_2015ApJ...798...44B}.

Another proposed mechanism is the irradiation feedback process, as suggested in \citet{Hameury_Ritter_1997A&AS..123..273H}, where a fraction of the released accretion luminosity is irradiated onto the photospheric layers of the companion star, primarily in the X-ray regime. 

Additionally, a portion of the pulsar wind particles can undergo reprocessing in the intrabinary shock and travel along the magnetic field lines of the companion star, reaching its surface. This mechanism, proposed to explain the heating of low-mass companions in black widow systems \citep{Sanchez_Romani_2017ApJ...845...42S,Yap_etal_2019A&A...621L...9Y}, involves the irradiation of the companion star. Furthermore, direct irradiation resulting from the shock region between the pulsar wind and the wind of the companion has been proposed. It should be noted, however, that this mechanism alone is unable to reproduce the observed modulation in the light curve \citep{Bogdanov_etal_2011ApJ...742...97B}. These various mechanisms can collectively contribute to the heating and increase in the surface temperature of the companion star.
In this work, we investigate the effect of heating the wind upstream of the companion by the pulsar wind by using the description given in \cite{tavani1993hydrodynamics}:

\begin{eqnarray}\label{Eq:heat_X_G_tavani_London_1993}
S_{\rm psr} &=& \rho \frac{\Lsd \sigma_T}{4 \pi a_{\rm IB}^2 m_p} \Upsilon \nonumber\\
&=& \rho \frac{\Lsd \sigma_T}{4 \pi a_{\rm IB}^2 m_p}\left [ \Upsilon_{\gamma} \left( 1- \frac{T}{T_c}\right) + \Upsilon_{x} \left( 1- \frac{T}{T_x}\right)\right]\,,
\end{eqnarray}
with $\Lsd$ the spin-down luminosity of the pulsar, $\sigma_T$ the Thomson cross section, and $m_p$ the proton mass. The heating parameters $\Upsilon$ is the measure of the quality of irradiating spectrum of X-rays and $\gamma$-rays, thus $\Upsilon_{\gamma}=(1-b)=0.1$ and $\Upsilon_{X}/\Upsilon_{\gamma} \ll 1$ as luminosity in the X-rays is assumed negligible. $T_c \sim 10^8 \rm K$ is the Compton temperature and $T_x \sim 5 \times 10^6 \rm K$ is the temperature at which the relevant heavy elements are almost completely ionized. These temperatures act as thresholds below which we consider the heating to be negligible.

This simplified approach cannot capture all relevant physics. Notably, the transferred matter through L1 may exhibit different physical properties, such as being in an optically thick regime. While we cannot account for such physics, we consider this approximation to be sufficient for simulating the interaction between the two winds at a significantly reduced computational cost compared to more comprehensive methods. 

We also introduce the power ratio, which relates the companion wind power to the fraction of pulsar spin-down power geometrically intercepted by the companion (assuming spherical symmetry of the pulsar wind):\begin{equation}\label{Eq:power_ratio}
\epsilon = \frac{\frac{1}{2}\mcdot \vc^2}{f\Lsd},
\end{equation}
where $f = R_{\rm c}/ (2\pi a_{\rm IB})$ is the fraction of power intercepted by the cross section of the companion in the equatorial plane with $R_{\rm c}$ the companion star radius, setting :
\begin{equation}
    f \simeq 0.05 \left(\frac{R_c}{0.3\times 10^{11}\rm cm}\right) \left(\frac{10^{11}\rm cm}{a_{\rm IB}}\right).
\end{equation}

\subsection{The cooling}\label{cooling}
We also incorporated an optically thin radiative cooling function $\Lambda(T)$ that depends on the local temperature $T$. We adopted the cooling function $\Lambda(T)$ described in the study by \citet{schure2009new}, which we calibrated for solar metallicity. Moreover, we maintained a floor temperature of 100 K.

\section{Intra-binary shock and X-ray emission}\label{sec:ibs_model}

\subsection{IBS analytical model}
The study of the interaction of two supersonics, isotropic stellar winds has been done in the past. 
Numerical simulations of the collision between two massive-star winds have been made using full gas-dynamics equations in the radiative regime in \citet{stevens1992colliding}. In this regime, the shocked region is thin and its locus can be brought down to the surface of ram-pressure balance as done in \citet{dyson1993mass}.

Analytical expressions of the shock surface between two spherical winds in the thin-shell regime have been discussed in \citet{canto1996exact}, as an extension of \citet{wilkin1996exact}.
These neglect the effect of orbital motion. The derivation rests upon conservation of mass, linear and angular momenta through the surface defined by ram-pressure balance. In this section we summarize the results of \citet{canto1996exact} that are relevant to our study.

The result is a surface of revolution around the intra-binary axis. The surface intersects the axis at the stagnation point located at a distance $r_{\rm sh}$ from the companion such that,
\begin{equation}
\label{eq:stag_radius}
    r_{\rm sh} = \frac{\beta^{1/2} a_{\rm IB} }{ 1 + \beta^{1/2}}\, ,
\end{equation}
where $\beta$ is the momentum flux ration defined in Eq. \ref{Eq:ratio_beta}. The surface is asymptotically conical at large distances. Its opening angle $\theta_\infty$ with respect to the companion-to-pulsar direction is given by the solution of :
\begin{equation}
    \theta_\infty - \tan \theta_\infty = \frac{\pi}{ 1 - \beta} \, .
\end{equation}
For example, when $\beta \rightarrow 0$ (pulsar wind infinitely dominating) we have $\theta_\infty\rightarrow \pi$, corresponding to a bow shock wrapping around the companion and asymptotically cylindrical.

These models offer a simple solution, providing an asymptotic angle for approximating the solid angle of Doppler emission within the shock. However, these models consistently neglect the influence of gravity and orbital motion. 

In the case of a close and thus fast-rotating binary system, such as a redback system, the Coriolis force is important as it breaks the cylindrical symmetry about the intra-binary axis, causing the stagnation surface to wrap and bend around the companion \citep{lemaster2007effect}. 

In this study, we perform hydrodynamical simulations of two isotropic winds, incorporating the effects of both gravity and orbital motion. 
It has been shown in \cite{bogovalov2012modelling} that a hydrodynamical approach provides a viable approximation for numerically modelling the interaction between a weakly magnetized pulsar wind and a stellar wind.

\subsection{IBS X-ray emission}\label{subsec:xray_emi}

The intrabinary-shock is an efficient site of particle acceleration and X-ray emission \citep{harding1990acceleration,arons1993high}.
The pulsar wind reaches relativistic bulk speeds along the shock, resulting in relativistic beaming of emissions by the shocked material. As a consequence, emissions from the shock are modulated with the orbital phase. 
Due to the beaming effect, the geometry of the shock surface and the inclination angle of the system largely determine the shape of the X-ray light curve. 
The X-ray flux is then showing a maximum at the inferior conjunction of the orbit for RBs (shock surrounding the pulsar), and a minimum for BWs (shock surrounding the companion). The single or double peak we can observe depends mainly on the opening angle of the intra-binary shock. 

In this study, we model the X-ray emission within the shock. For this purpose we evaluate the SR of an electron population in the IBS and solve the radiative transfer equation along different lines of sight, allowing us to study the link between the geometry of the shock and the associated light curve in X-rays. 

Taking physics variables from our simulations allows us to differ from previous analytical works. First of all the direction of particle propagation no longer strictly adheres to the shock control line, but instead aligns with the velocity of the flux within the IBS. Secondly the emission in the shock depending on the electron population is modeled using the density in the simulation. 

We consider a relativistic electron population in each shock cell $i$, set with a power law as :
\begin{equation}
    N_i(\gamma_e) d\gamma_e = N_{0,i}\gamma_e^{-p}d\gamma_e \,,
\end{equation}

with $\gamma_{\rm e, min} < \gamma_e < \gamma_{\rm e, max}$. 
The constant $p$ depends on the process of particle acceleration. 
X-ray spectra of tMSPs show a hard spectrum corresponding to a particle distribution index $p \sim 1.3$ \citep[][and references therein]{papitto2021transitional}. 
As already noted in \citep[e.g.,][]{kandel2019synchrotron}, a value close to 1 suggests that the dominant mechanism of particle acceleration is magnetic reconnection \citep{sironi2014relativistic}. 
However, determining a dominant acceleration process based solely on a hard spectrum is challenging, especially when considering the spectrum of injected particles from the pulsar wind \citep{van2020x}. As a result, the acceleration process is not considered in this work. For simplicity, we adopt the fiducial value p = 1.3. 
$N_0$ is the normalization coefficient, defined as \citep{gomez1995parsec}:

\begin{equation}
    N_{0} = \left [ \frac{e_{th,e}(p-2)}{1-\gamma_R^{2-p}}\right ] ^{p-1} \left [ \frac{1-\gamma_R^{1-p}}{n_{e}(p-1)} \right ]^{p-2}  \,,
\end{equation}
where $e_{th,e} = \epsilon_e e_{th}$ and $n_{e} = \epsilon_e n$. $e_{th}$ is here the thermal energy, $n$ the number density and $\epsilon_e=0.1$ represents the fraction of electrons contribution. We set $\gamma_R=\gamma_{\rm e, max}/\gamma_{\rm e, min}$, and $\gamma_{\rm e}$ is obtained via the observed frequency of synchrotron in X-ray band,  $\nu_{\rm min-max}= 2.4 \times 10^{16}-2.4 \times 10^{18}$ Hz (corresponding to the range of energy $0.1-10$keV \citep{bogdanov2009chandra,huang2012x}) :

\begin{equation}
   \gamma_{\rm e, min-max} = \sqrt{\frac{2\pi m_e c \nu_{\rm min-max}}{e B}} \, .
\end{equation}

If we consider the magnetic field along the shock is defined (assuming a split monopole geometry \citep{thompson2004magnetar,weber1967angular}) toroidal outside the light cylinder as :

\begin{equation}
    B(r_{\rm sh}) = B_{\rm LC}\left(\frac{r_{\rm LC}}{r_{\rm sh}}\right)
,\end{equation}
with
\begin{equation}
    B_{\rm LC}=B_{\rm surf}\left(\frac{r_{\rm surf}}{r_{\rm LC}}\right)^3 \,.
\end{equation}
Using the typical values for MSPs ($P=10^{-3}s$ , $\Dot{P}=10^{-20}s.s^{-1}$, $r_{\rm surf}=10$km), we obtain $B_{\rm surf}\sim10^{8} $G. With $r_{\rm LC}=c/\Omega=50$km, we have $B_{\rm LC}\sim 8\times 10^{5} $G. For each simulation we compute emission assuming a magnetic field at the stagnation point of $B_{\rm sh} = 80$ G, meaning $\gamma_e\sim 10^{4}-10^{5}$.

We also consider the radiative losses to be instantaneous and the synchrotron emissivity occurring in the electron rest-frame as :

\begin{equation}
    j_{\gamma_e}(\nu)=\frac{P_{\gamma}(\nu)}{4\pi}=\frac{\sqrt{3}e^3Bsin(\alpha)}{4\pi m_e c^2}\left( \frac{\nu}{\nu_c}\right)\int_{\nu/\nu_c}^{\infty}K_{5/3}(x)dx
    \label{gamma cell rf} \,,
\end{equation}
with
\begin{equation}
    \nu_c = 3/2 \gamma_e^2 \frac{eBsin(\alpha)}{2\pi m_e c} \,.
\end{equation}
We assume that particles move in a circular motion around the magnetic field lines in the restframe, meaning that the pitch angle $\alpha$ is 90°.
The total specific emission in each cell is then computed as :
\begin{equation}
    j(\nu) = \int d\gamma_e N(\gamma_e) j_{\gamma_e}(\nu) \,.
\label{total_cell_rf}
\end{equation}

In order to obtain the specific intensity received by an observer, the quantity of Eq. \ref{total_cell_rf} is transformed, 
\begin{equation}
    j_{\nu}' = \delta^{2} j_{\nu}\,,
\end{equation}
where $\delta$ is the Doppler factor :
\begin{equation}
    \delta = \frac{1}{\Gamma(1-\sqrt{1-1/\Gamma^2}\cos(\theta))}\,,
\end{equation}
with $\Gamma$ the bulk Lorentz factor of pulsar material, $\theta$ the angle between the flow direction and sky position from the considered line of sight. The flow direction is directly computed using the velocity of the matter in the simulation.

The orbital variability of the total emission is obtained by solving the radiative transfer equation along a line of sight for a different viewing angles during all the orbital period, following :
\begin{equation}
\label{Inu}
 I_{\nu} = I_{\nu,0}\exp{(-\tau_{\nu})} + \int_0^{\tau_{\nu}} \delta^{2} j_{\nu}\exp{-(\tau_{\nu}}-\tau_{\nu}\prime) d\tau_{\nu}\prime \,.
\end{equation} 
We consider the optical depth $\tau_{\nu}\ll 1$.

Since we are in 2D, the stars always eclipse the line of sight. The radius of the stellar companion being far more predominant than the neutron star (NS), we consider the emission to be blocked only if the star is in the LOS. 
The associated X-ray light curve of IBS occurring in our simulations are computed to constrain our model.

\section{Simulation setup}\label{sec:simulations_setup}
The simulations were conducted using the Message Passing Interface-Adaptive Mesh Refinement Versatile Advection Code (MPI-AMRVAC; \citealt{keppens2021mpi}). We incorporated the treatment of a rotating frame around the center of mass, which can be shifted from the grid center, into the hydrodynamic module of the AMRVAC code.

In this modified version of AMRVAC, we solved the conservation equations for mass, momentum, and energy. The effects of the gravity of both stars and the rotation of the reference frame were included as source terms in the momentum and energy equations. Additionally, heating and optically thin radiative cooling were accounted for through source terms in the energy equation. The following equations were solved:

\begin{eqnarray}
\frac{\partial \rho}{\partial t} + \vec{\nabla} \cdot (\rho \vec{v} ) &= &0\,,\\
\frac{\partial \rho \vec{v}}{\partial t} + \vec{\nabla} \cdot (\rho \vec{v} \vec{v} + P)& =& -\rho \vec{\nabla}\Phi_{\rm g} - 2 \rho \vec{\Omega} \times \vec{v} \nonumber\\
&&- \rho \vec{\Omega} \times \left[\vec{\Omega} \times \vec{r}_{\rm center}\right]\,,\\
\frac{\partial e}{\partial t} + \nabla \cdot \left[\left(e+P\right) \vec{v} \right] &=& - \rho \vec{v}\cdot\vec{\nabla}\Phi_{\rm g} -\left(\frac{\rho}{m_{h}}\right)^2\Lambda(T)+ S_{\rm psr} \nonumber\\
&&- \rho \vec{v} \cdot \left[ \vec{\Omega} \times \left(\vec{\Omega} \times \vec{r}_{\rm center}\right) \right]\,,
\end{eqnarray}
where $\rho$ and $P$ represent the fluid density and pressure, respectively; $\vec{v}$ denotes the fluid velocity vector in the reference frame. The total energy in the co-rotating frame is given by $e = e_{\rm th} + \frac{1}{2} \rho v^2$, where $e_{\rm th}$ represents the thermal energy. In this paper, we assume a polytropic equation of state such that $e_{\rm th}=\frac{p}{\gamma-1}$ and $\gamma=5/3$ the polytropic index. After conducting some tests, we did not include viscosity in the above equations because it did not have a significant effect over the simulation timescales.

The gravitational potential is defined as $\Phi_{\rm g}=-\frac{G M_{\rm c}}{r_{\rm c}}-\frac{G M_{\rm psr}}{r_{\rm psr}}$, where $r_{\rm c}$ represents the distance from the companion star and $r_{\rm psr}$ represents the distance from the pulsar.
Regarding the rotating reference frame, it has an angular velocity $\vec{\Omega}$ centered on the central mass, and within the grid, the distance from this center is denoted as $r_{\rm center}$.
Regarding the heating, our study primarily investigates the impact of X-ray and $\gamma$-ray irradiation from the pulsar on the companion wind. This is represented by the source term  $S_{\rm psr}$ (Eq. \ref{Eq:heat_X_G_tavani_London_1993}).
The cooling term $\Lambda(T)$ is  discussed in Sect. ~\ref{cooling}.

The simulations are conducted on a 2D polar grid, covering a radial extent of $r=[3\times10^{9}, 2\times10^{11}]$ cm and a toroidal extent of $\phi=[0,2\pi]$ (see Fig . \ref{sketch}). A logarithmic grid is employed in the radial direction. Initially, the grid is refined to the first level with a resolution of $(32\times64)$ cells. Given the substantial scale difference between the shock scales, the pulsar wind zone, and the overall simulation box investigated in this project, the use of adaptive mesh refinement (AMR) is essential to accurately resolve the significant variations encountered in the spider system with a pulsar. In our simulations, the grid is allowed to be refined up to 9 additional levels, with a doubling of resolution at each new level of refinement. After studying the impact of resolution, we determined that this level of resolution is the minimum necessary to adequately capture the physics close to the pulsar. The refinement and coarsening processes utilize Lohner's error estimator to assess second-order variations in density and radial velocity. Moreover, the refinement is only allowed in regions where the winds are encountered.

Regarding the internal and external boundary conditions, the companion star is placed at the grid center, and its wind is imposed within the equipotential surface with an effective potential $\Phi_{\rm c,surface}$. 
Concerning the boundary condition around the NS, during the accretion phase study, an inflow Neumann boundary condition projected on the radial direction in the pulsar's frame is applied. The radial outer boundaries are treated as open boundaries, utilizing a Neumann boundary condition with vanishing derivatives for all quantities.
During the active pulsar phase the wind is emitted from the surface of a sphere with a radius of $R_{\rm psr,w}$. 
The equations are solved using the total variation diminishing (TVD) method, specifically the Harten-Lax-van Leer-Contact (HLLC) method \citep{harten1997high}, combined with the Koren flux limiter \citep{vreugdenhil1993numerical}. 

\section{Simulations of characteristics states and transition}\label{sec:results}

In the context of transitional systems, there is an intricate progression from a LMXB state, characterized by discernible accretion onto the NS, to a pulsar state where radio emissions remain unobscured \citet{stappers2014state}. 
Within this section, our principal aim is to replicate both of these states through the utilization of our computational model. Subsequently, we focus on the influence of the companion wind characteristics on the state of the system and the corresponding observables. 

Here, we study different stellar wind configurations, each with a different effect on the behavior of the system. Our overarching objective is to delineate the critical threshold that marks the transition from a pulsar accreting state to a radio pulsar state.

\subsection{Accretion stream without a pulsar wind}\label{subsec:1w}

First, we set the parameters for an accreting neutron star and the wind from the companion star. Specifically, the wind of the NS is inactive and the companion wind is adjusted to ensure that mass transfer occurs by RLOF.
Table \ref{tab:stellar_wind_parameters} presents the matrix of cases we have studied. Rows are labeled from "0" through "3" in order of increasing companion wind speed $v_{\rm c,\, w}$, while columns labeled "a" and "b" are in order of increasing companion mass flux, $\dot{m}_{\rm c, w}$.

\begin{table}
    \centering
    \caption{Mass flux and speed of the companion star wind, and the corresponding power ratio (assuming a pulsar spin-down luminosity $\Lsd = 10^{35}$ erg/s).}
    \begin{tabular}{|c|c|c|}
        \hline
         $\dot{m}_{\text{c,w}}\, (\text{M}_{\odot}/\text{yr})$ & \makecell{(a) \\  $4.5\times10^{-10}$ } & \makecell{(b) \\  $9\times10^{-10}$}  \\
        \hline
        \hline
        \makecell{(0)\\  $v_{\rm c,\, w} = 60$ km/s\\ $\xi_{\rm c, rot}\sim 0.12$ \\  $\xi_{\rm c, grav}\sim 0.4$} & \makecell{$\epsilon=1.13\times10^{-4}$} & \makecell{$\epsilon=2.26\times10^{-4}$  } \\
        \hline
        \makecell{(1)\\  $v_{\rm c,\, w} = 100$ km/s \\ $\xi_{\rm c, rot}\sim 0.2$ \\  $\xi_{\rm c, grav}\sim 0.67$} & \makecell{$\epsilon=3.14\times10^{-4}$ } & \makecell{$\epsilon=6.29\times10^{-4}$ } \\ 
        \hline
        \makecell{(2)\\  $v_{\rm c,\, w} = 150$ km/s  \\ $\xi_{\rm c, rot}\sim 0.3$ \\  $\xi_{\rm c, grav}\sim 1$ } & \makecell{$\epsilon=7.08\times10^{-4}$ } & \makecell{$\epsilon=1.41\times10^{-3}$ } \\ 
        \hline
        \makecell{(3)\\   $v_{\rm c,\, w} = 200$ km/s  \\ $\xi_{\rm c, rot}\sim 0.4$ \\ $ \xi_{\rm c, grav}\sim 1.33$} & \makecell{$\epsilon=1.25\times10^{-3}$ } & \makecell{$\epsilon=2.51\times10^{-3}$ } \\ 
        \hline
    \end{tabular}
    \tablefoot{For a stellar mechanical luminosity $L_{\rm c}=1\times10^{30}$ erg/s, we have $\epsilon\sim 2 \times 10^{-4}$. The influence of gravity and orbital motion are characterized by ratios Eqs. \ref{Eq:vrot_ratio} and \ref{Eq:vmin_ratio}.}
    \label{tab:stellar_wind_parameters}
\end{table}

\begin{figure}
\begin{center}
\includegraphics[width=\columnwidth]{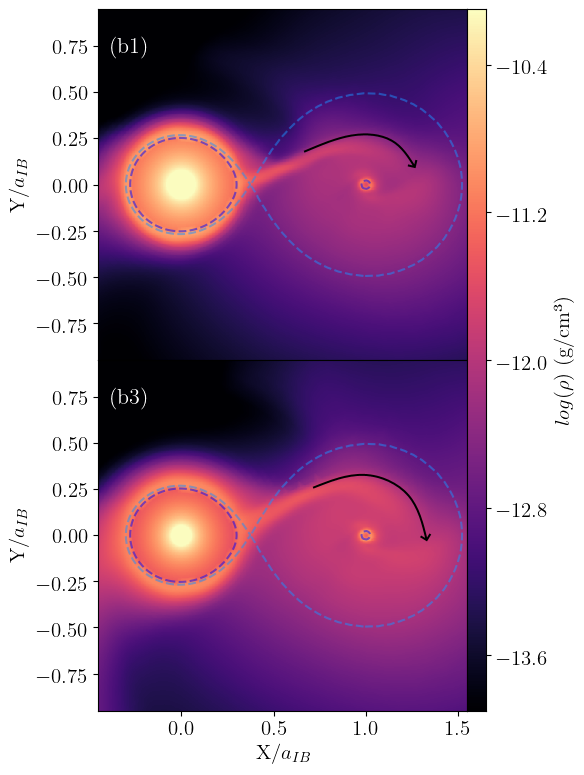}
\caption{Comparison of the accretion stream in cases (b1) and (b3) without pulsar wind, that is for two different companion wind velocities and equal companion mass loss. Left is the companion and right is the NS. Purple dashed lines indicate boundary surfaces, while the blue dashed line indicates the isocontour of the effective gravitational potential through the L1 Lagrange point. The black arrow shows the direction of orbital rotation. The color scale gives matter density.}
\label{Fig:arm}
\end{center}
\end{figure}

Due to the effects of gravity and orbital motion, the supersonic accretion stream leaving $\rm L_1$ does not directly collide with the surface of the pulsar but rather orbit it eccentrically.
Then if the dynamics is dominated by the gravity of the neutron star, the angular momentum of the stream with respect to the neutron star $l_{\rm stream}$ gives the circularization radius following \citep{hendriks2023mass} : 

\begin{equation}\label{Eq:Keplerian_circularization_radius}
    R_{\rm circ} = \frac{l_{\rm stream}^2}{G\,M_{\rm psr}} \,.
\end{equation}

This leads to a resulting circularization radius of $ R_{\rm circ} \approx (1-x_{1})^4\,a_{\rm IB}\,\frac{1+q}{q}\approx 0.2\aib$, with $q=M_{\rm psr}/M_{\rm c}=3.5$ the mass ratio of the system and $x_{1}\sim 0.37$ the distance of $\rm L_1$ from the companion star \citep{lu2023rapid}.  
In our simulations we obtain radii within a range $R_{\rm circ,b0}\sim0.92 R_{\rm circ}$ and $R_{\rm circ,b3}\sim1.35R_{\rm circ}$ ($R_{\rm circ,b1}\sim0.98R_{\rm circ}$ and $R_{\rm circ,b2}\sim1.1R_{\rm circ}$).
In fact, as the speed of the companion wind increases, the angular momentum of the accretion stream increases, so that it circulates at a greater distance from the neutron star.

Due to the rotation of the system, the returning stream then collides with itself (the incident stream). This interaction redirects the subsequent paths of the incident stream and the returning stream following the rotation direction as illustrated in Fig. \ref{Fig:arm}. 
The impact does not disrupt the upstream trajectory of the incident stream since no disturbance can propagate upstream in supersonic flows. However, each new collision induces additional deflections that result in the circularization of the flow, following the dynamics explained in \cite{lubow1975gas}. 
Within these interacting streams, a residual disk of material forms from earlier interactions, with the material slowly spiraling inward toward the NS.

The variability in the mean density enveloping the pulsar underscores the interaction between the returning stream and the incident stream (Fig. \ref{fig:cc_changes}, left column). 
Within a ring delimited by the internal boundary of the pulsar wind and the estimated circularization radius $R_{\rm circ}$, two types of mass loss are observed. 
The first one, occurring twice per period, is associated with an angular momentum variation, indicating changes in the circularization radius over time as the stream approaches and then moves away from the control ring. 
The second one is manifesting approximately 10 times per orbital period which corresponds to the Keplerian period at the circularization radius. 
Notably, in the case of (b3), while the second mass loss persists, the first one is less pronounced, indicating a stationary accretion stream state. This can be attributed to the increased density within the flow, resulting in less perturbation of the accretion stream. This stationary behavior is particularly pertinent when examining the interaction with the pulsar wind.

The increase in companion wind speed leads to an increase in the specific kinetic energy and thus Bernoulli energy within the accretion stream. This results in a lateral expansion of the stream over a wider equipotential surface in the vicinity of $\rm L_1$. Specifically, the width of the accretion stream is 3 times larger in case (b1) than in case (b0), and 3.2 times larger in case (b3) than in case (b0). 
Additionally, examining cases (a0) to (a3) reveals that the thickness remains consistent with cases (b0) to (b3), respectively (from 0.048 $\aib$ to 0.288 $\aib$), and that this parameter is only influenced by the speed.

Another consequence of this phenomenon is the enhanced ram pressure within the accretion stream. The total ram pressure in the accretion stream shows a direct correlation with the initial speed of the stellar wind.

Near $\rm L_1$, in the inertial frame, the kinetic energy of the accretion stream is comparable to the thermal energy, with Mach numbers $\mathcal{M}_{\rm L1} < 1$ (between $\mathcal{M}_{\rm L1,b0}=0.7$ and $\mathcal{M}_{\rm L1,b3}=0.3$).
As the stream approaches the neutron star, it is accelerated, with the acceleration being more pronounced for scenario (b0) compared to scenario (b3). Consequently, the transition to the supersonic regime occurs earlier at lower speeds.
As the flow is accelerated along the stream, the energy becomes mainly kinetic when approaching the NS, accounting for more than 90\% of the energy for (b0) and 70\% for (b3).

Within the accretion stream, the Bernoulli energy (Eq. \ref{Eq:companion_wind_bernoulie}) remains constant along the streamlines, indicating an adiabatic flow.
However, the thickness of the stream decreases under tidal effect, while the mass flux stays constant, causing the overflowing stream to become more confined near the stream center. 

Regarding mass loss in the system, apart from the matter not accreted by the NS and ejected in the close environment, we observe a condensed mass loss phenomenon at the $\rm L_2$ point behind the companion star. This differs from the perspective proposed by \cite{lu2023rapid}, which focuses on potential mass loss of an accretion disk at the $\rm L_3$ point. 
However, it is noteworthy that the total mass lost remains negligible when compared to the mass efficiently captured by the NS gravitational field.

\subsection{Pulsar wind influence on accretion stream}\label{subsec:2w}

\begin{figure*}
\begin{center}
\includegraphics[width=0.97\textwidth]{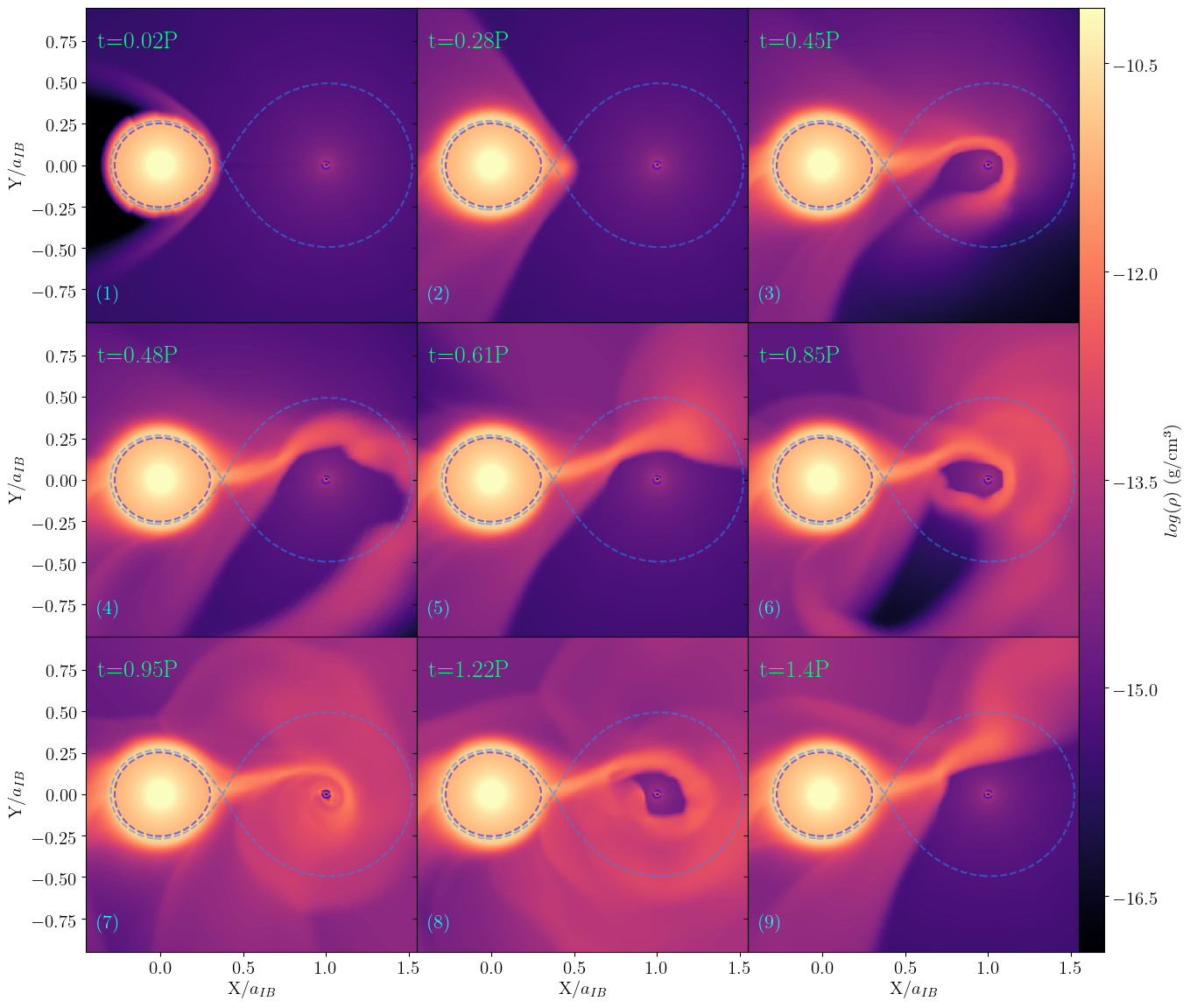}
\caption{Visualization of the interaction between the accretion stream and pulsar wind for the case (b1). Left is the companion and right is the NS. Purple dashed lines indicate boundary surfaces, while the blue dashed line indicates the isocontour of the effective gravitational potential through the L1 Lagrange point. The color scale gives matter density. Green text top left indicates the time in unit of orbital period, $P$.
}
\label{Fig:state_evol}
\end{center}
\end{figure*}

The pulsar wind is modeled as described in Sect. \ref{psr wind} and the settings are summarized in Table \ref{tab:pulsar_parameters}.
Figure (\ref{Fig:state_evol}) shows the evolution of the scenario (b1), which serves as a representative example of all simulations involving the interaction between the accretion stream and the pulsar wind. These cases are represented by orange cells in Table \ref{tab:ratios}.

The fast pulsar wind reaches the Roche Lobe of the companion star and a shock forms around the companion star (see panel (1)).
If the ram pressure within the stream is sufficiently high, the accretion stream emerges (panel $(2)$) and reaches the NS (panel $(3)$). 
The circularization radius remains the same across the different stellar parameters and is smaller compared to inactive pulsar wind scenarios (Sect. \ref{subsec:1w}), with $R_{\rm circ,b1}\sim R_{\rm circ,b3} \sim0.8R_{\rm circ}$. The influence of the pulsar wind leads to a decrease in the velocity of the companion wind, which in turn leads to a decrease in the specific angular momentum and a smaller circularization radius. 
The interaction with the pulsar wind also causes the stream center to be shifted from the $\rm L_1$ point by approximately 0.05 $\aib$ in the direction orthogonal to the line of centers (visible in panel $(7)$). 
On top of that we observe that the transition to a supersonic regime is now occuring precisely at $\rm L_1$, $\mathcal{M}_{\rm L1} \sim 1$.

During the third phase, the fast pulsar wind interacts strongly with the accretion stream. It compresses it laterally, pushing it further away from the pulsar. The closest distance between the accretion stream and the pulsar varies between $0.1-0.2 \aib$ (panels $(4)$ and $(5)$). 
During this variable phase, the accretion stream continues to wrap around the pulsar until the head of the accretion stream completes its tour around the pulsar and intersects with the side of the incoming stream at about $0.85P$.
At the same time, the stream rotates and converges toward the pulsar and confines the pulsar wind (panel $(6)$).
The system then enters an accretion stream regime, akin to what was observed with an isolated companion wind, the pulsar wind now mixed in the accretion flow surrounding the NS (panel $(7)$).

\begin{figure*}
\includegraphics[scale=0.58]{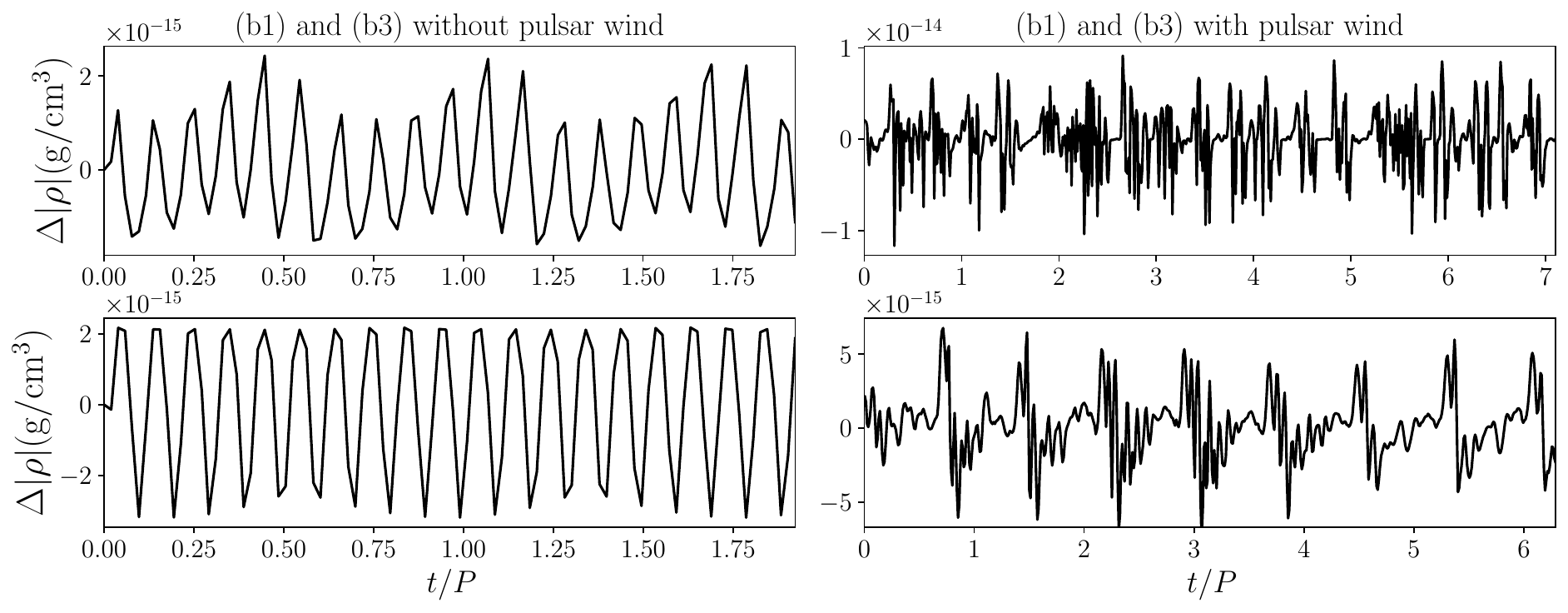}
    \caption{ Variability of the accretion stream regime with and without pulsar wind. The panels show, from top to bottom, the temporal variation of mean density (in $\rm g.cm^{-3}$) within a ring, extending from the internal boundary of the pulsar wind to the circularization radius $R_{\rm circ}$. Left column correspond to simulations without a pulsar wind (see Sect. \ref{subsec:1w}), while right column to simulation with both winds. First line corresponds to scenario (b1), second line to scenario (b3). 
    }
    \label{fig:cc_changes}
\end{figure*}

Unlike the periodic behavior we used to observe considering an isolated companion wind (see Sect. \ref{subsec:1w}), the accretion stream state, when interacting with a pulsar wind, exhibits increased instability.
We now witness seemingly chaotic behaviour on timescales of a fraction of orbital period (Fig. \ref{fig:cc_changes}, right column). 
Due to those instabilities, the ram pressure of the pulsar wind close to the NS on occasion surpass that of the accretion stream. 
In the case of the weak accretion stream (b1),  around time $\rm t\sim 1.4P$, the pulsar wind significantly disturbs the accretion arm, pushing it outward. This process returns the system to a state similar to the initial transition phase at time $\rm t\sim 0.8P$ (as seen in panels (8) and (9)).
This initiates an unstable behavior in which the accretion stream is expelled several times until the returning stream successfully catches up with the incident one and the system returns to an accretion stream state akin to panel (7) (at about $\rm t\sim 5P$), before undergoing disruption once again. This instability highlights the intricate interaction between the two winds.

However, scenarios (b2) and (b3) exhibit a more stable accretion stream state, remaining relatively unaffected by the pulsar wind, thereby maintaining the configuration shown in panel $(7)$, Fig. \ref{Fig:state_evol}.
The system then enters into a quasi-stationary state, where instabilities quasi-periodically lead to the accretion stream being disrupted away before gradually rebuilding (Fig. \ref{fig:cc_changes}, bottom right panel). However the matter surrounding the NS is not entirely expelled, maintaining the system in a accretion stream state.
These instabilities occurring around the pulsar are intricately tied to the accumulation of matter that remains unaccreted. This accumulation leads to an increase in the total angular momentum, reaching a point where it becomes too high to sustain a stable flow.
Consequently, mass expulsion occurs followed by renewed accumulation, effectively restoring the system to its previous density.
Therefore the rate at which these instabilities manifest is directly related to the pace at which matter accumulates. Faster accumulation results in more frequent instabilities.
In Sect. \ref{subsec:1w}, we demonstrated that increasing the initial speed of the companion wind leads to an increase in density within the stream. This results in an increase in the frequency of instabilities. 
Indeed in our simulations, instabilities occur approximately every $0.8P$ for scenario (b3) and approximately every $1P$ for scenario (b2).
It is noteworthy that each instability is distinct, even though the system exhibits a repeating pattern resembling a limit cycle.

A similar variability is also observed in PSR J1023+0038 while in the LMXB state, as it notably exhibits a distinctive bimodal behavior in X-ray luminosity \citep{papitto2015x,bogdanov2014x,archibald2015accretion,de2013x}, characterized by low and high modes. 
The high mode prevails approximately 80\% of the time, characterized by a luminosity of approximately $3 \times 10^{33}$ erg/s. In contrast, the low mode occurs about 20\% of the time, and is characterized by a luminosity around $5 \times 10^{32}$ erg/s. X-ray flares, detected occasionally, exhibit luminosities reaching up to $5 \times 10^{34}$ erg/s but are observed in less than 2\% of the time.
However, distinct from our simulations where instabilities occur on the order of hours, transitions between the low and high states occur rapidly, with a timescale of approximately 10 seconds, and no periodicity seems to emerge \citep{tendulkar2014nustar,bogdanov2015coordinated}.

The mechanism governing transitions between the high and low states remains elusive and seems reliant on inhomogeneities in the accretion flow.
The discovery of this bimodal flux distribution \citep{bogdanov2015coordinated,patruno2013new} triggered discussions regarding possible accretion in the propeller regime with matter reaching the magnetic poles of the NS for the high mode \citep{Papitto_Torres_2015ApJ...807...33P,campana2016physical,romanova2005propeller}.
The low mode on the other hand suggests shock emission from the interaction of the relativistic pulsar wind with the accretion disk \citep{linares2014x,zelati2018simultaneous}, surviving just outside the light cylinder \citep{eksi2004disks} and advancing back and forth toward the corotation radius.
However the corotating radius and light cylinder, which are in our case approximately $R_{\rm co}\approx16.7$ km and $R_{\rm LC}\approx50$ km, remains unresolved in our simulation, and we can only suggest a potential link between these two seemingly chaotic variabilities. Moreover, we cannot speculate on a possible IR/optical signature of the behavior observed in our model based on our current considerations.

\subsection{Radio pulsar state and transition}\label{subsec:transition}

\begin{figure}
\begin{center}
\includegraphics[width=\columnwidth]{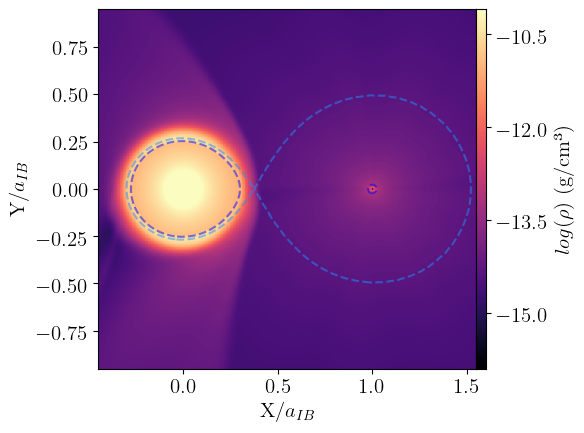}
\caption{Stationary "PSR" state outcome for case (b0) obtained in case of a dominant pulsar wind. Left is the companion and right is the NS. Purple dashed lines indicate boundary surfaces, while the blue dashed line indicates the isocontour of the effective gravitational potential through the L1 Lagrange point.  The color scale gives matter density.}
\label{Fig:state}
\end{center}
\end{figure}

Depending on the mechanical luminosity ratio between the companion wind and the pulsar wind, there are cases where the pulsar wind dominates, potentially preventing the formation of the accretion stream from the companion, as in cases (a0), (b0), and (a1) (Fig. \ref{Fig:state}). Conversely, there are certain cases where the companion wind is strong enough to give rise to an accretion stream, even in cases where its mechanical luminosity is lower than that of the pulsar, as observed in the above cases (panel (7) Fig. \ref{Fig:state_evol}). The occurrence of these different regimes can be described by the mechanical luminosity ratio between the companion and pulsar winds, as listed in Table \ref{tab:ratios}.
The transition between these states is explained by examining the ram pressure in both the accretion stream and the pulsar wind. 

Indeed as explained Sect.\ref{sec:ibs_model}, wind-wind interaction is usually characterized by the momentum flux ratio $\betaiso$ (Eq.~\ref{Eq:ratio_beta}), determining the shape of the stagnation surface between the two winds but also its precise location $\rsh$(Eq. \ref{eq:stag_radius}) \citep{stevens1992colliding, johnstone2015colliding}. 
However, this ratio describes the entire competition between two isotropic winds when neglecting the effects of the Coriolis force and gravity. 
In our scenario, where gravity and orbital motion significantly influence wind interaction, it becomes crucial to analyze the ram pressure's evolution within the accretion stream. 
Indeed in the context of the RLOF process, the kinetic energy, ram pressure and density within the stream passing through $\rm L_1$ surpasses that of an isotropic wind. This distinction arises from the condensed and confined flow, channeled through a single point.  
As a result, there is no smooth transition from a pulsar-dominated to a companion-dominated wind as in the isotropic case.

\begin{figure}[h]
    \centering
    \includegraphics[width=\columnwidth]{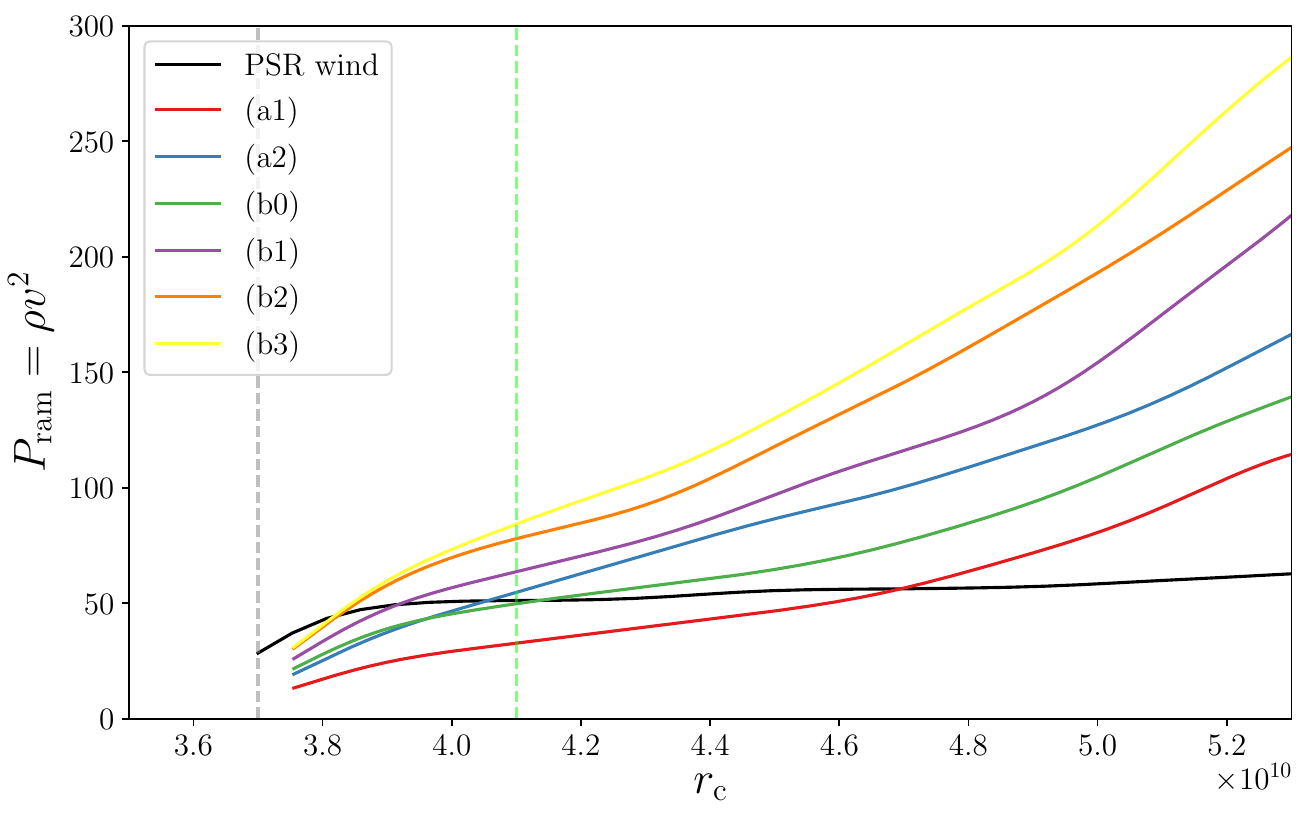}
    \caption{Evolution of the ram pressure  as a function of the distance from the companion star. The colored curves correspond to the accretion stream (for simulations without a pulsar wind), while the black curve corresponds to the pulsar wind.The critical radius $\rcrit$ is shown here with the vertical green line. The vertical gray line indicates the position of $\rm L_1$.}
    \label{fig:pram}
\end{figure}

The radius $\rt$ is the stagnation position from the companion star, where the accretion stream and the pulsar wind are balanced, such that the ram pressures are equal, $\rho_{\rm c,\,w}(\rt)\,v^2_{\rm c,\,w}(\rt) = \rho_{\rm psr,\,w}(\rt)\,v^2_{\rm psr,\,w}(\rt)$. 
This radius for each cases investigated is shown in Fig. \ref{fig:pram} represented by the different intersections between the curve of the pulsar wind and accretion stream ram pressure. 

We have established the existence of a critical transition radius $r_{\rm t,crit}$ such that for $\rt < r_{\rm t,crit}$ an accretion stream is formed but not for $\rt > r_{\rm t,crit}$. In the latter case, the system enters a radio pulsar state characterized by an intra-binary shock wrapping around the companion star. Its stagnation radius lies before the critical radius and an embryonic arm shields the companion from the pulsar wind.

In the isotropic-wind picture the transition between pulsar-dominated and companion dominated occurs at $\betaiso = 1$ with an IBS symetrically wrapping around one or the other component. Here, due to the effect of gravity, the settings required for the companion wind to become dominant and create an accretion stream are far less demanding in energy. We thus propose a more insightful approach by computing the ratio of ram pressure between the pulsar wind and the accretion stream (in simulations without a pulsar wind) at the critical radius $r_{\rm t,crit}$.
This ratio is defined as :
\begin{equation}\label{Eq:ratio_eta}
    \eta = \frac{\rho_{\rm c,\,w}(r_{\rm t,crit})\,v^2_{\rm c,\,w}(r_{\rm t,crit}) }{\rho_{\rm psr,\,w}(r_{\rm t,crit})\,v^2_{\rm psr,\,w}(r_{\rm t,crit})} \,.
\end{equation}
By construction, the transition occurs when $\eta = 1$. On the other hand whenever $\eta < 1$, indicating that the pulsar wind dominates the ram pressure at $r_{\rm t,crit}$, the system is in a pulsar state. Conversely, a dominant companion wind at $r_{\rm t,crit}$, $\eta >1$, leads to the formation of an accretion stream. The transition is occurring for relatively low value of $\betaiso$ compared to analytical works (see Table \ref{tab:ratios}).

We observe a critical radius at $r_{\rm t,crit}\sim0.41 \aib$, consistent with the instances where the accretion stream does not appear (scenarios (b0) and (a1) Fig. \ref{fig:pram}). 
The time required for the stream to fully emerge is prolonged when $r_t \rightarrow r_{\rm t,crit}^-$. 

The distinctive discontinuity in the dominance of the winds underscores the contrast to isotropic systems due to gravity, orbital motion, and the RLOF regime.
The existence of this tipping point might provide a plausible explanation for the transitional phenomena observed in spider systems, suggesting that these systems may be approaching a configuration close to this critical limit. 
However, we note that the instabilities observed in our simulations occur over relatively short timescales (a few hours) compared to the state transitions observed in tMSPs, which occur over several years \citep{stappers2014state}.

\begin{table}
\caption{Outcome of the system depending on wind parameters and ratios $\betaiso$ and $\eta$.}
\begin{NiceTabular}{|c|c|c|}
    \hline
    $\dot{m}_{\rm c,w} (\mathrm{M_{\odot}/yr})$ & \makecell{(a)  \hspace{0.5cm}  $4.5\times10^{-10}$  } & \makecell{(b)  \hspace{0.5cm} $9\times10^{-10}$ }  \\
    \hline
    \hline
    \makecell{(0)\\  $v_{\rm c,\, w} = 60$ km/s} & \cellcolor{blue!50}\makecell{$\betaiso= 0.06$ , \\ $\eta= 0.51$} & \cellcolor{blue!50}\makecell{$\betaiso= 0.12$ , \\ $\eta= 0.97$  } \\
    \hline
    \makecell{(1)\\  $v_{\rm c,\, w} = 100$ km/s } & \cellcolor{blue!50}\makecell{$\betaiso= 0.1$, \\ $\eta= 0.64$ } & \cellcolor{orange!50}\makecell{$\betaiso= 0.2$, \\$\eta= 1.24$ } \\ 
    \hline
    \makecell{(2)\\  $v_{\rm c,\, w} = 150$ km/s  } & \cellcolor{orange!50}\makecell{$\betaiso= 0.15$, \\$\eta= 1.07$ } & \cellcolor{orange!50}\makecell{$\betaiso= 0.3$, \\ $\eta= 1.53$ } \\ 
    \hline
    \makecell{(3)\\   $v_{\rm c,\, w} = 200$ km/s} & \cellcolor{orange!50}\makecell{$\betaiso= 0.2$, \\ $\eta= 1.23$ } & \cellcolor{orange!50}\makecell{$\betaiso= 0.4$, \\ $\eta= 1.65$ } \\ 
    \hline
\end{NiceTabular}
\tablefoot{ Orange cells correspond to "ACC" state shown Fig. \ref{Fig:state_evol}, blue to "PSR" state shown Fig. \ref{Fig:state}. }
\label{tab:ratios}
\end{table}


\section{Orbital variability of the X-ray flux in the pulsar state}\label{sec:xLC}

In this section we apply the model of X-ray emissions exposed in Sec. \ref{subsec:xray_emi} to the pulsar state in order to produce light curves. We set aside the accretion stream state since no stationary IBS forms and therefore the model does not apply.

\subsection{Shock detection}\label{subsec:lfact}
We use the shock detection approach proposed by \cite{lehmann2016shockfind}. First, the candidate shock cells are identified using the norm of the relative density gradient,
\begin{equation}
     |\nabla \rho | = \sqrt{(\partial_x \rho)^2+(\partial_y \rho)^2+(\partial_z \rho)^2} \,,
\end{equation}
and the convergence of velocity using :

\begin{equation}
     - \nabla \cdot \mathbf{u}  = - (\partial_x u_x+\partial_y u_y+\partial_z u_z) \,.
\end{equation}

Candidate cells are identified based on a density gradient or a velocity convergence surpassing a user-defined threshold specific to each simulation.
This approach enables us to selectively focus on the termination front of the pulsar wind, typically corresponding to the region of the shock that is of particular interest in our study.\\

\subsection{Analogous relativistic bulk velocity}

The emission model requires a relativistic bulk velocity within the shock, particularly to account for relativistic beaming. We prescribe that the relativistic specific momentum be proportional to the simulated non-relativistic one, that is,
\begin{equation}
\frac{\Gamma \vec{v}_{\rm R}}{\Gamma_{\rm max} c} = \frac{\vec{v}_{\rm sim}}{v_{\rm sim, max}},
\end{equation}
where $ \vec{v}_{\rm R}$ is the relativistic velocity, $\Gamma$ the corresponding Lorentz factor, $\vec{v}_{\rm sim}$ the simulated velocity, and $v_{\rm sim, max}$ the maximum detected velocity within the shock. In practice the latter corresponds to the asymptotic velocity along the arm. We follow \citet{wadiasingh2017constraining,kandel2019synchrotron} in order to set the asymptotic Lorentz factor to $\Gamma_{\rm max} = 1.8$.

\subsection{X-ray light curves}\label{subsec:LC}
\begin{figure*}%
    \centering
    \includegraphics[width=0.97\textwidth]{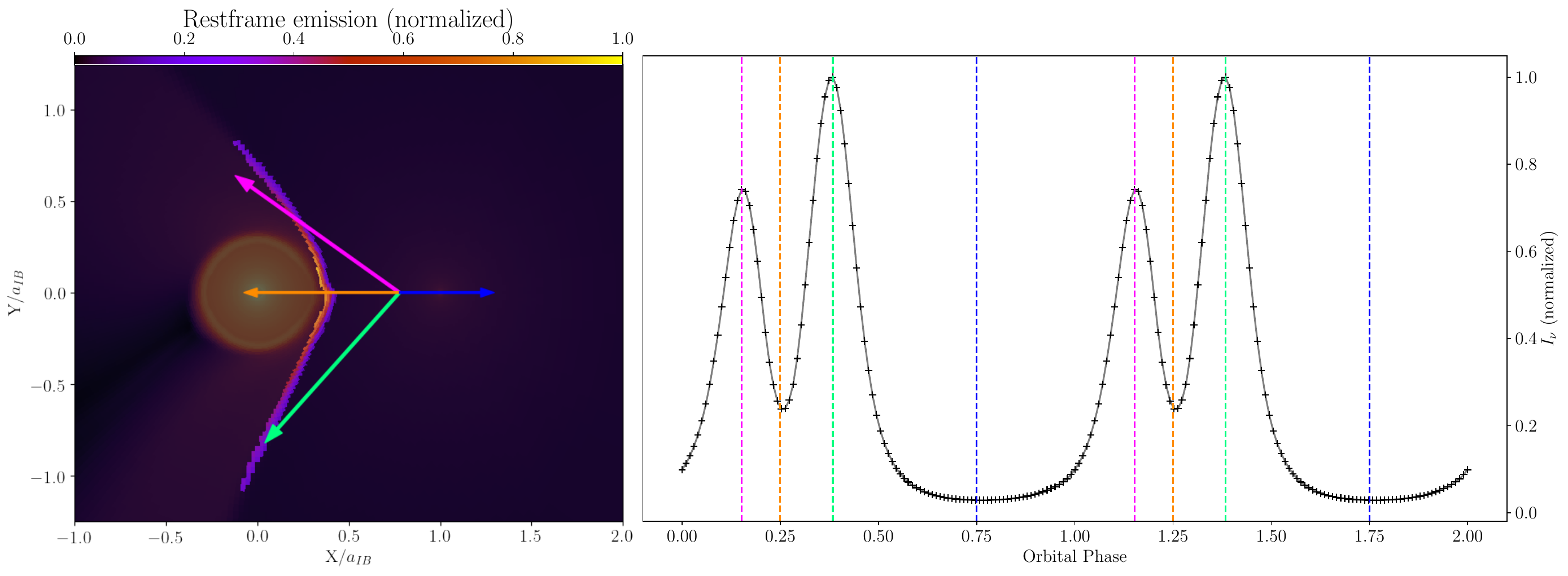}
    \caption{Left panel: Detected shocks cells for the pulsar radio state (b0) (also see Fig. \ref{Fig:state}), with the color scale representing the intensity of the rest-frame synchrotron emission normalized to its maximum value. Arrows indicate the lines of sight corresponding to the vectical lines with the same colors on the right panel. Right panel: Associated X-ray light curve. The orange and blue dashed lines correspond respectively to the star and pulsar inferior conjunctions (at 0.25 and 0.75 of the orbital phase). The vertical magenta and green dashed lines correspond to the angles with the maximum intensity received from the system.}%
    \label{Fig:shock_LC}%
\end{figure*}

The shock and light curve in Fig. \ref{Fig:shock_LC} correspond to case (b0) and are representative of the three cases in the pulsar state (a0,  a1, b0).
The light curve has a periodic double-peak feature around the companion star inferior conjunction, aligning with expectations for an IBS created due to a dominating pulsar wind.
In analytical studies, the directions where the maximum emission occurs typically corresponds to the asymptotic direction of the shock, as explained in Sect. \ref{sec:ibs_model}.

However in our case we note that it does not correspond exactly to this direction.  In our simulation the flow crosses the shock surface at a non-negligible angle which mostly explains the observed difference. This allows the regions closer to the apex of the shock, where the rest-frame emission is strongest, to make a larger contribution to the total flux. Consequently, the maximum emission, resulting from Doppler boosting, is achieved closer to the star's inferior conjunction. 
Additionally, an eclipse in the emission at precisely 0.25 of the orbital phase further accentuates the double-peak pattern, resulting from obstruction by the companion star itself.

Furthermore the double-peak feature is not perfectly centered at the star's inferior conjunction (green and magenta arrows on left panel Fig. \ref{Fig:shock_LC}). 
Indeed, due to the orbital motion, the shock is not entirely symmetrical when surrounding the companion. 

In addition, we observe an amplitude asymmetry of about 20\% between the peaks: the leading peak has a lower luminosity compared to the trailing one (see right panel of Fig. \ref{Fig:shock_LC}). 
Indeed, part of the emission contributing to both peaks originates from the apex region of the shock, where a significant fraction is eclipsed by the companion star. 
As a result of orbital motion, the position of the apex region is shifted and the eclipse is asymmetric. This accounts for about half of the amplitude difference. 
Another effect is that particles are relatively more accelerated in the trailing arm of the shock, accounting for the remaining amplitude difference.  
These effects should be all the more pronounced that the orbit is viewed edge-on, and we therefore propose that such asymmetry is an indication of large inclination of the system.

The emission observed near the pulsar inferior conjunction corresponds to the apex of the shock surface where the plasma can be considered at rest and emits isotropically without being eclipsed. This represents approximately 10\% of the highest emission.

Comparing our generated light curve with known systems, we can draw parallels with PSR B1957+20 \citep{huang2012x,kandel2021xmm}, which has a double-peak pattern around the star's inferior conjunction. Despite the differences in stellar properties between our simulations and this system, we expect a similar light curve shape due to its edge-on inclination \cite{clark2023neutron}.
In \citet{kandel2021xmm}, we can clearly see a pronounced eclipse between the two peaks, where the minimum flux even falls below the flux at pulsar inferior conjunction. This suggests that the shock may be closer to the companion star or more relativistically beamed than in our simulation. Additionally, the asymmetry  between the two peaks of the light curve of PSR B1957+20 is consistent with that of Fig. \ref{Fig:shock_LC}, with the leading peak being smaller and the trailing one being larger. This supports the idea that such an asymmetry should be regarded as an indication of the system being nearly edge-on.

PSR J2129-0429 is an example of a RB system where the X-ray light curve is interpreted by an IBS wrapping around the pulsar instead of the companion \citep{roberts2015x,al2018x,kong2018broad}.  
Indeed, a double peak is observed around inferior conjunction of the pulsar. No asymmetry is observed between the amplitudes of the two peaks which is consistent with the fact that the pulsar cannot eclipse the shock. For the same reason, the dip between the two peaks can only be due to beaming in this model.
A shock surrounding the pulsar is usually observed in RB cases and obtained in studies where the gravity and orbital motion are not factored in. In such studies, an isotropic companion wind with dominating ram pressure results in the bending of the shock toward the pulsar \citep{Romani_Sanchez_2016ApJ...828....7R,wadiasingh2018pressure,kandel2019synchrotron}.
As explained in Sect. \ref{subsec:1w} the configuration of the stellar wind in our simulation is set to create an accretion stream through RLOF. As a result the IBS created arises from the impact with the accretion stream rather than an isotropic companion wind. Instead, a situation similar to the aforementioned studies would require the companion wind speed to be much greater than the system's gravitational escape velocity, thereby preventing RLOF.
\\

We also note different effects of the momentum flux ratio $\betaiso$ on the obtained light curve.
The difference between the angles at which the maximum of emission occurs, corresponding to the two peaks, is linked to the opening angle of the shock. This angle diminishes as the ratio $\betaiso$ decreases. This phenomenon arises from the fact that a powerful pulsar wind generates a shock that progressively surrounds the companion, resulting in a less distinct double-peak feature.
The stagnation radius of the shock created is also strongly influenced  by the momentum fluxes ratio. It decreases from $r_{\rm sh} = 0.4 a_{\rm IB}$ close to the critical radius, to $r_{\rm sh}\sim 0.3 a_{\rm IB}$ as the shock will directly hit the surface of the companion for $\betaiso \ll 1$.

\normalfont
\color{black}

\section{Conclusions}
We employed the AMRVAC 2.0 code to perform high-precision 2D HD simulations of the interaction between the winds of a stellar companion and a pulsar in a RB system. The goal of our study has been to create a model taking into account the effects of gravity, orbital motion, and heating from the pulsar wind. 

To this end, we created a simplified model for both winds.  Numerous approximations have been made in prior studies and this work is meant as a step toward a more self-consistent treatment of the complex phenomenology in these rich systems.
The pulsar wind is non-relativistic, but settings were selected to reproduce an interaction with a typical MSP of $\Lsd=10^{35}$erg/s, achieved by simulating a wind with an equivalent momentum flux. 
For the companion wind, we explored different values of initial speed and mass flux in order to characterize the process of accretion through RLOF in a spider binary.

This approach allowed us to shed light on the influence of the parameters of the companion wind on the shape and density of the resulting accretion stream.
We replicated the two characteristic states observed in transitional systems: the accretion stream state and the radio pulsar state.

In contrast to analytical works, we observed that the transition between these two states is not gradual but instead undergoes a sharp shift at a specific tipping point.
The transition radius, $\rt$, marks when the ram pressure of the accretion stream becomes dominant over the pulsar wind. We defined the critical radius, $r_{\rm t,crit}$, such that for $\rt<r_{\rm t,crit}$, the stream emerges and reaches the NS. Otherwise the pulsar wind dominates and the system enters a radio pulsar state characterized by an IBS encircling the companion star. 
We note that the transition occurs for momentum flux ratios $\beta \sim 0.14$. This is much smaller than the value of 1 at which the system becomes dominated by the companion wind in analytical works that neglect gravity. This underscores the importance of incorporating gravity and orbital motion in our model.

Furthermore, we witnessed the instabilities induced by the pulsar wind while in the accretion stream state. In proximity to the tipping point, we identified an unstable behavior characterized by the intermittent creation and disruption of the accretion stream state. 
Although these alternations of pulsar and radio phases occur on timescales of a few orbital periods in our simulations (i.e., on a much shorter timescale than the few years observed for tMSPs), we suggest that this hints at a plausible explanation for the transitional phenomena observed. Indeed, these systems could be in a state close to a similar tipping point.

The IBS was chosen as the observable for comparison with known observations. 
We constructed the corresponding X-ray light curves modeling the SR emission occurring in the shock for the radio pulsar state. This state corresponds to a BW scenario where the pulsar wind dominates. The RB case, where the shock is surrounding the pulsar, was not stable with the physical parameters used in this work.
The orbital motion shifts the position of the peaks with respect to the inferior conjunction of the companion. Notably, the leading peak is less intense than the trailing peak due to both the orbital motion and the occultation of the shock front apex by the companion. We suggest that such an asymmetry could be indicative of high inclination. Indeed, a similar asymmetry is observed in the light curve of the system PSR B1957+20, which is almost edge-on.

Our 2D approach confines us to the orbital plane. A 3D model would allow us to verify if the accretion flow remains within the orbital plane or if the expulsion of matter by the pulsar wind occurs in a preferred direction. Additionally, 3D modeling would enable examination of the impact of the system's inclination on the observables. This would allow for observations of the shock morphology due to the interaction with the accretion flow in the direction perpendicular to the orbital plane.

The follow-up to this study will be to consider a ray tracing algorithm for the heating model. 
The ideas to consider a relativistic wind for the pulsar and doing MHD simulation to take into account the magnetic field impact on the transitions are also really important to constrain our system as much as possible.

\begin{acknowledgement}
We would like to express our gratitude to the referee for their insightful and pertinent questions, which significantly contributed to improving the quality of this paper.
\end{acknowledgement}
\def\bibfont{\footnotesize}
\bibliographystyle{aa}
\typeout{}
\bibliography{aa50638_24.bib}   
\addcontentsline{toc}{section}{Bibliographie}

\end{document}